\documentclass[a4paper,11pt]{article}
\pdfoutput=1 
\usepackage{jheppub} 
\usepackage{mathrsfs}
\usepackage{amsmath,amsfonts,latexsym,amssymb,graphics,epsfig,subfigure,color,makeidx}
\usepackage{xcolor}
\usepackage{epstopdf}
\usepackage{epsfig,subfigure}
\newcommand {\nn}{\nonumber}
\newcommand {\pt}{\partial}

\title{Localization of q-form fields on a de Sitter brane in chameleon gravity}
\author[a,b]{Yi Zhong,} 
\author[c,*]{Tao-Tao Sui \note[*]{Corresponding authors}}
\affiliation[a]{School of Physics and Electronics Science,
             Hunan University, Changsha 410082, China}
\affiliation[b]{Lanzhou Center for Theoretical Physics, Key Laboratory of Theoretical Physics of Gansu Province, and Key Laboratory of Quantum Theory and Applications of MoE, Lanzhou University, Lanzhou, Gansu 730000, China}
\affiliation[c]{College of Physics, Nanjing University of Aeronautics and Astronautics, Nanjing 211106, China}

\emailAdd{zhongy@hnu.edu.cn}
\emailAdd{suitt14@lzu.edu.cn}

\abstract{Recently, it was found that the vector field can be naturally localized on the thick brane in  chameleon gravity. In this work, we extend this study to encompass de Sitter brane scenario. We focus on the localization of q-form fields. The scalar and vector fields can be localized on the de Sitter brane, while the KR field cannot be localized. The condition for localization of the scalar and vector fields is obtained. Furthermore, we investigate the localization characteristics in two examples with given conformal factor $b(\phi)$. In the first case,  the effective potentials and KK modes of the matter fields are obtained asymmetric even though the de Sitter brane has $Z_2$ symmetry. In the second case, volcano-like effective potentials are generated in the de Sitter brane model.}

\begin{document} 
\maketitle

\section{Introduction}\label{scheme1}
The idea that there may exist an extra spatial dimension has been an interesting topic ever since the proposal of the Kaluza-Klein (KK) theory in the 1920s. An intriguing theory within this field is the Randall-Sundrum (RS) model, which offers a potential solution to a perplexing puzzle in physics known as the hierarchy problem \cite{Randall1999,Randall1999a}. This problem revolves around the question of why gravity seems weaker compared to the other fundamental  forces. In the RS model, the brane is a four-dimensional hypersurface embedded in a five-dimensional spacetime known as the "bulk." However, due to the geometrically thin nature of the brane, the curvature becomes divergent on the brane. Hence, the concept of the thick brane or domain wall is introduced, wherein the domain wall is a topological defect of the  background scalar field in the bulk \cite{Gremm2000b,Gass1999,Gremm2000,Afonso2006}. In the braneworld scenario, gravity is free
to propagate in the bulk, while the matter fields  are confined on the brane. As a result, the essential role of the localization mechanism for matter fields emerges within the braneworld theory's framework. This mechanism not only yields a four-dimensional effective theory at low energy scales but also provides potential effects of the extra dimension \cite{Bajc2000,Oda2000,Melfo2006,Liu2009,Almeida2009,Liang2009,Vaquera-Araujo2015,Salvio2007}. 
In other words, the action of matter fields in five-dimensional spacetime can be dimensionally reduced to a zero-mass four-dimensional action along with a series of massive four-dimensional actions. The zero-mass four-dimensional action corresponds to particles in the Standard Model, while the massive four-dimensional actions correspond to excited KK particles at high energy scales.
The free scalar fields can be localized on the branes , while the localization of fermion fields requires the  Yukawa couplings with background scalar field \cite{Liu2018b}. The localization of the q-form fields has been investigated in Refs.\cite{Fu2012,Fu2014,Fu2016}. However, the localization mechanism of the vector field remains an open question. Some researchers have sought to address this by introducing couplings between the vector field and the background scalar field, coupling between the vector field and spacetime curvature to tackle the issue of vector field localization \cite{Vaquera-Araujo2015,Chumbes2012a,Belchior2023b}.

On the other hand, general relativity has been found to encounter several challenges, encompassing both phenomenological and theoretical issues. Among these challenges are the dark energy problem, and the singularity problem. In order to address these intricate issues, numerous alternative theories of gravity have been put forth. And the braneworld theory has also been extended to the framework of modified gravities and new features have been found \cite{Liu2018c}. For example, it was found that thick branes can be generated by pure geometry in Weyl gravity and $f(R)$ gravity \cite{Barbosa-Cendejas2006b,Zhong2016}, and may have inner structure in mimetic gravity and $f(T)$ gravity \cite{Zhong2018,Yang2012}. 

Recently, it was found that the vector field can be naturally localized on the thick brane in the chameleon gravity \cite{Zhong2023}. The chameleon gravity was proposed to explain the dark energy problem \cite{Khoury2004,Khoury2004a}. In this theory, matter fields couple to a physical metric  $\widetilde{g}_{\mu\nu}$, and  $\widetilde{g}_{\mu\nu}$ is related to the metric $g_{\mu\nu}$ which describe the spacetime geometry by  a conformal transformation with a function of the chameleon scalar field $\phi$, i.e. $\widetilde{g}_{\mu\nu}=b^2(\phi)g_{\mu\nu}$. 
The chameleon scalar field varies in mass depending on the density of surrounding matter. This environmental nature of the screening means that such a scalar field may be relevant on other scales \cite{Burrage2018}. The highly-nonlinear nature of screening mechanisms means that they evade classical fifth-force searches, and there has been an intense effort towards designing new and novel tests to probe them, both in the laboratory and using astrophysical objects, and by reinterpreting existing datasets \cite{Burrage2018}. A similar screen mechanism has been studied in $f(R)$ gravity, and this type of $f(R)$ gravity is mathematically equivalent to chameleon gravity, where the function $f(R)$  corresponds to the conformal factor $b(\phi)$ in chameleon gravity \cite{DeFelice2010,Burrage2018}.

In chameleon gravity, various matter fields couple to the chameleon scalar field $\phi$ through the physical metric  $\widetilde{g}_{\mu\nu}$, and this provides a potential mechanism to localize the vector field. Brane cosmology in chameleon gravity was investigated in Refs.~\cite{Saaidi2012,Bisabr2017}. In Ref. \cite{Zhong2023}, localization of various matter fields on a flat brane was investigated and new features were found. Considering the accelerated expansion of our universe, there is a compelling incentive to extend this study to encompass the de Sitter brane scenario. See Refs. \cite{Liu2009a,Liu2009b,Guerrero2019,Yang2020,Guo2013} for works on the localization of matter fields on a de Sitter brane. In this work, we will focus on the localization of q-form fields on a de Sitter brane in chameleon gravity. 

The layout of the paper is as follows: In Section 2, we construct a  de Sitter brane in the framework of chameleon gravity. In Section 3, we investigate the localization of q-form fields on the de Sitter brane. In Section 4, we provide the specific form of the conformal factor $b(\phi)$ and investigate two cases with novel localization properties. 
Throughout this paper, $M,N,\cdots $ denote the indices of the five-dimensional coordinates, and $\mu,\nu,\cdots$ denote the ones on the brane.

\section{De Sitter brane in chameleon gravity}
In this study, we begin with the following five-dimensional action which describes a canonical scalar field minimally coupled to gravity \cite{Burrage2018a}:
\begin{equation}
 \label{action1}
S=\int d^5 x \sqrt{-g}\left[ \frac{1}{2}R-\frac{1}{2}\nabla_M \phi \nabla^M \phi -V(\phi) \right]+ S_m [\tilde{g}_{MN},\psi_i],
\end{equation}
where the  matter fields $\psi_i$ coupled to the Jordan frame metric $\tilde{g}_{MN}=b^2 (\phi) g_{MN}$. 
We will not specify the specific form of the conformal factor initially; instead, we will engage in a general discussion. Later, in Section 4, we will discuss its specific form.

The metric of a de-Sitter brane is assumed as
    \begin{eqnarray}
        \label{metric}
       ds^2 &=& \text{e}^{2A(z)}\left(-dt^2+\text{e}^{2Ht}(dx_1^2+dx_2^2+dx_3^2 )+dz^2 \right)\nn\\
        &=& \text{e}^{2A(z)}\left(\gamma_{\mu\nu}dx^{\mu}dx^{\nu}+dz^2 \right) ,
    \end{eqnarray}
where $z$ stands for the extra coordina, $H$ is the Hubble parameter, and $H>0$.  $a(z)=\text{e}^{A(z)}$ is the warp factor, $\gamma_{\mu\nu}dx^{\mu}dx^{\nu}=-dt^2+\text{e}^{2Ht}(dx_1^2+dx_2^2+dx_3^2 )$.
Since the brane is maximally symmetric, the scalar field can be assumed as $\phi=\phi(z)$. When constructing the braneworld background solution, we neglect the influence of matter fields on the brane, i.e., we set $S_m=0$.
With the metric assumption of (\ref{metric}), and varying the action (\ref{action1}) with respect to the Einstein frame metric $g_{MN}$ and the scalar field $\phi$, we obtain the following equations
    \begin{eqnarray}
        \label{eom1}
        3A'^2-\phi'^2-3A''-3 H^2 &=&0, \\
        \label{eom2}
        2\text{e}^{2A}V(\phi)+ 9(A'^2- H^2)+3A''
                 &=&0,	\\
        \label{eom3}
      	 4A'\left(\phi'^2+6 H^2\right)-\text{e}^{2A}\phi' V'(\phi )&=&0,
    \end{eqnarray}
where the primes denote the derivatives with respect to $z$.
Note that there are only two independent equations in Eqs. (\ref{eom1})-(\ref{eom3}),
We assume the following warp factor of a de sitter brane \cite{Kobayashi2002a},
    \begin{eqnarray}
    	\label{wf2}
 	\text{e}^{2A(z)} = \text{sech}^{p}(k z),
    \end{eqnarray}
 with the scalar field solved as
    \begin{eqnarray}
 	\phi(z) = \phi_0 \text{arcsin}\left(\tanh(k z)\right),
    \end{eqnarray}
where $\phi_0=\left[3p(1-p)\right]^\frac{1}{2}$, $k=\frac{H}{p}$ and $0<p<1$.
The corresponding potential of the scalar field is
       \begin{eqnarray}
 	V(\phi) = \frac{1}{2} k ^2 p \left(9 p+3\right) \cos ^{2 (1-p)}\left(\frac{\phi }{\phi _0}\right).
    \end{eqnarray}

    \begin{figure}[!htb]
    \begin{center}
    \subfigure[The warp factor]{\includegraphics[width=6cm]{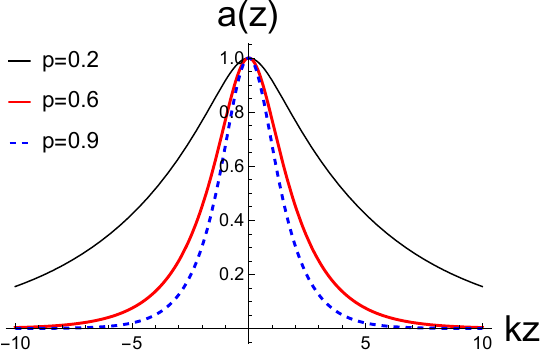}}
    \subfigure[The scalar field]{\includegraphics[width=6cm]{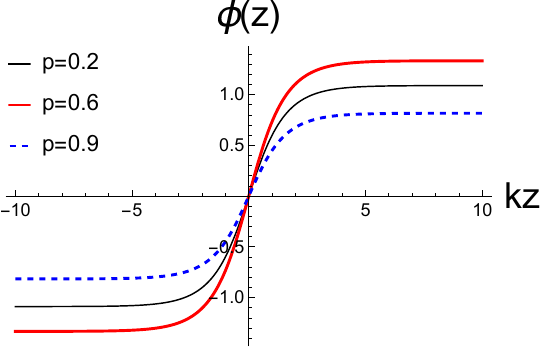}}
    \end{center}
    \caption{The shapes of the warp factor $a(z)$ and the scalar field $\phi(z)$ in the de Sitter brane model.}\label{flat}
    \end{figure}
For convenience, we express the Jordan frame metric as follows:
\begin{equation}
\label{tilde g}
\tilde{ds}^2=\tilde{g}_{MN}dx^{M}dx^{N}=\tilde{a}^2(z)
\left[ \gamma_{\mu\nu}dx^{\mu}dx^{\nu}+dz^2
\right],
\end{equation}
where we introduce an effective warped factor $\tilde{a}(z)=\text{e}^{\tilde{A}(z)}=\text{e}^{B(\phi)+A(z)}$, and $b(\phi)=\text{e}^{B(\phi)}$. All functions derived from $\tilde {g}_{MN}$ will be denoted with a tilde in the following. Notably, the braneworld solution in this setup remains the same as the standard case. 
Moreover, since the four-dimensional gravity is only related to the tensor mode of the perturbation of the metric,  the localization of gravity in our model is exactly the same with that in the standard case in general relativity, and it is not affected by the conformal factor $b(\phi)$.
However, we anticipate novel characteristics concerning the localization of various matter fields since the matter fields couple to the Jordan frame metric $\tilde{g}_{MN}$ instead of the Einstein frame metric $g_{MN}$.

\section{Localization of the q-form fields}
Firstly, we focus on localizing the $q$-form field $X_{M_1 M_2 \ldots M_q}$ which corresponds to a scalar field when $q=0$, a vector field when $q=1$, and a Kalb-Ramond (KR) field when $q=2$ on a de-Sitter brane. Our approach involves investigating a general $q$-form field first and then delving into the specific cases of scalar, vector, and Kalb-Ramond fields.
The research regarding the localization of $q$-form fields on a flat brane in general relativity can be found in Refs.~\cite{Fu2012,Fu2014,Fu2016}. Similar derivations of the formulas of a q-form field in a de sitter brane model can be found in Ref. \cite{Yang2020}.
The action of a massless $q$-form field in five-dimension is given by
\begin{equation}
\label{q-form action}
    S_q = -\frac{1}{2(q+1)!} \int d^5 x \sqrt{-\tilde{g}} Y^{M_1 M_2 \ldots M_{q+1}} Y_{M_1 M_2 \ldots M_{q+1}},
\end{equation}
where $\tilde{g}$ represents the determinant of the Jordan frame metric $\tilde{g}_{MN}$. The indices of the matter fields are raised and lowered using $\tilde{g}_{MN}$, and the strength of the $q$-form field is defined as $Y_{M_1 M_2 \ldots M_{q+1}}=\partial_{[M_1} X_{M_2 M_3 \ldots M_{q+1}]}$.
By varying action (\ref{q-form action}) with respect to $X_{M_1 M_2 \ldots M_q}$, the equations of motion of the $q$-form field can be obtained,
\begin{eqnarray}
\label{q-form eom0}
       \pt_{M_1} \left(\sqrt{-\tilde{g}} Y^{M_1 M_2 \ldots M_{q+1}} \right)  =0.
\end{eqnarray}

By substituting the Jordan frame metric (\ref{tilde g}), we can get the equations of motion as
\begin{eqnarray}
\label{q-form eom1}
       \pt_{\mu_1} \left(\sqrt{-\tilde{g}} Y^{\mu_1 \mu_2 \ldots \mu_{q+1}} \right) + \pt_{z} \left(\sqrt{-\tilde{g}} Y^{z \mu_2 \ldots \mu_{q+1}} \right) &=&0,\\
    \label{q-form eom2}
        \pt_{\mu_1} \left(\sqrt{-\tilde{g}} Y^{\mu_1 \ldots \mu_{q} z} \right)&=&0.
\end{eqnarray}
We should note that the five-dimension massless $q$-form field action (\ref{q-form action}) is invariant under the gauge transformation $X_{M_1 M_2 \ldots M_q} \rightarrow \tilde{X}_{M_1 M_2 \ldots M_q}=X_{M_1 M_2 \ldots M_q}+\pt_{[M_1} \Lambda_{M_2 M_3 \ldots M_{q}]}$, where $\Lambda_{M_2 M_3 \ldots M_{q}}$ can be considered as an arbitrary antisymmetric tensor field. In order to remove the normative degrees of freedom, we can choose the gauge $X_{\mu_1 \ldots \mu_{q-1} z}=0$.  Then, the rest components of the $q$-form field with the KK decomposition can be expressed as
\begin{eqnarray}
\label{q-form kk}
    X_{\mu_1 \mu_2\ldots \mu_{q}}(x, z)= \sum_{n} \hat{X}_{\mu_1 \mu_2\ldots \mu_{q}}^{(n)}(x)\Upsilon_n(z)\text{e}^{\frac{(2q-3)\tilde{A}}{2}}.
\end{eqnarray}
Here, `n' can be interpreted as representing different KK modes, and as a result, we can obtain the following KK decomposition for the strength
\begin{eqnarray}
\label{q-form kk2}
Y_{\mu_1 \mu_2 \ldots \mu_{q+1}}(x, z) &=& \sum_{n} \hat{Y}_{\mu_1 \mu_2 \ldots \mu_{q+1}}^{(n)}(x)\Upsilon_n(z) \text{e}^{\frac{(2q-3)\tilde{A}}{2}},   \nn\\
Y_{\mu_1 \mu_2 \ldots \mu_q z}(x, z) &=& \frac{1}{q+1}\sum_{n} \hat{X}_{\mu_1 \ldots \mu_{q}}(x)  \text{e}^{\frac{(2q-3)\tilde{A}}{2}}\nonumber\\
&\times& \left(\Upsilon'_{n}(z)+\frac{(2q-3)}{2} \tilde{A}'\Upsilon_{n}(z) \right),
\end{eqnarray}
where $\hat{Y}_{\mu_1 \mu_2 \ldots \mu_{q+1}}^{(n)}(x)=\partial_{[\mu 1} \hat{X}_{\mu_2 \ldots \mu_{q+1}]}(x)$ is the field strength on the brane.  
With the above KK decomposition, and substituting the relations Eqs. \eqref{q-form kk2} into the equations of motion \eqref{q-form eom1}, we can reduce the equations of motion \eqref{q-form eom1} into 
\begin{eqnarray}\label{q-form eom3}
&&\sum_{n}\Big\{\partial_{\mu_1} \big(\sqrt{-\tilde{g}} \hat{Y}^{\mu_1 \mu_2 \ldots \mu_{q+1}}_{(n)}(x)\big)\Upsilon_n(z)\text{e}^{\frac{(2q-3)\tilde{A}}{2}} \\
&&+\frac{1}{q+1}\partial_{z} \Big(\sqrt{-\tilde{g}}\partial_z\big(\Upsilon_n(z)\text{e}^{\frac{(2q-3)\tilde{A}}{2}}\big)\Big)\hat{X}^{\mu_2 \mu_3\ldots \mu_{q+1}}_{(n)}(x)\Big\}=0.\nn
\end{eqnarray}
For different KK modes, we can see that they are independent from each other. In other words, we can discuss the different KK modes separately. So, for the `n-th' KK mode, we divide the left and right sides of above equation by $\hat{X}^{\mu_2 \mu_3\ldots \mu_{q+1}}_{(n)}(x)\Upsilon_n(z)\text{e}^{\frac{(2q-3)\tilde{A}}{2}}\sqrt{-\gamma}$ at the same time, 
and note that $\sqrt{-\tilde{g}}=\text{e}^{5\tilde{A}}\sqrt{-\gamma}$, then we obtain 
\begin{eqnarray}\label{q-form eom3a}
\frac{\partial_{\mu_1} \big(\sqrt{-\gamma} \hat{Y}^{\mu_1 \mu_2 \ldots \mu_{q+1}}_{(n)}(x)\big)}{\sqrt{-\gamma} \hat{X}^{\mu_2 \mu_3\ldots \mu_{q+1}}_{(n)}(x)}
+\frac{1}{q+1}\frac{\partial_{z}\Big(\text{e}^{5\tilde{A}}\partial_z\big(\Upsilon_n(z)\text{e}^{\frac{(2q-3)\tilde{A}}{2}}\big)\Big)}{\Upsilon_n(z)\text{e}^{\frac{(2q-3)\tilde{A}}{2}}}=0.
\end{eqnarray}
We can see that equation \eqref{q-form eom3a} is made up of two parts, the first term is a function of the four-dimensional coordinates, and the second term is of the extra dimension. Therefore, both the two terms must vanish. The first term of  the four-dimensional part $\hat{Y}_{\mu_1 \mu_2 \ldots \mu_{q+1}}^{(n)}(x)$  
\begin{eqnarray}\label{q-form 4d}
\pt_{\mu_1}\left(\sqrt{-\gamma}\hat{Y}^{\mu_1 \mu_2 \ldots \mu_{q+1}}_{(n)}(x)\right)
=\frac{1}{q+1}m_n^2 X^{\mu_2 \ldots \mu_{q+1}}\sqrt{-\gamma},
\end{eqnarray}
where $m_n$ is the mass of the KK mode, 
and second part yields 
\begin{eqnarray}\label{q-form eom3b}
 \partial_{z}\Big(\text{e}^{5\tilde{A}}\partial_z\big(\Upsilon_n(z)\text{e}^{\frac{(2q-3)\tilde{A}}{2}}\big)\Big)=
m_n^2 \Upsilon_n(z)\text{e}^{\frac{(2q-3)\tilde{A}}{2}}.
\end{eqnarray}
 extra dimensional which can be considered as a Schr\"odinger-like equation of the  KK modes $\Upsilon_n(z)$,
 \begin{eqnarray}
\label{q-form schro}
    \left[ -\pt^2  _z +V_q (z) \right] \Upsilon_n(z)=m_n ^2 \Upsilon_n(z),
\end{eqnarray}	
where the $m_n$ can be considered as the effective mass of the four-dimensional q-form field, and the effective potential $V_q (z)$ is
  \begin{eqnarray}
\label{q-form potential}
    V_q (z) = \frac{3-2q}{2}\tilde{A}'' + \frac{(3-2q)^2}{4} \tilde{A}'^2.
\end{eqnarray}
The derivation of the Schr\"odinger-like equation above is an extension of the derivation process of the Schr\"odinger-like equation in Ref.~\cite{Zhong2023}, where the flat four-dimensional metric is replaced with an de Sitter metric $\gamma_{\mu\nu}$ that is independent of $z$.
In order to ensure that the five-dimensional action (\ref{q-form action}) reduces to a four-dimensional effective action on the brane, the integral of the extra part should be convergent,  which can be realized by introducing the orthonormal condition
\begin{eqnarray}
\label{orth condition}
    \int^{+\infty}_{-\infty} \Upsilon_{m}(z) \Upsilon_{n}(z) dz = \delta_{mn}.
\end{eqnarray}
The action (\ref{q-form action}) can be reduced to
\begin{eqnarray}
\label{q-form eff action}
 S_q=&&-\frac{1}{2(q+1)} \sum_{n} \int d^4 x \sqrt{-\gamma}  \Big(\hat{Y}_{\mu_1 \mu_2 \ldots \mu_{q+1}}^{(n)} \hat{Y}^{(n) \mu_1 \mu_2 \ldots \mu_{q+1}} \! \nn\\
 &&+\frac{m_n^2}{q+1}\hat{X}_{\mu_1 \mu_2 \ldots \mu_{q}}^{(n)} \hat{X}^{(n) \mu_1 \mu_2 \ldots \mu_{q}} \Big) ,
\end{eqnarray}
where the indices are raised and lowered by the  metric $\gamma_{\mu\nu}$.  The  effective action describes a massless $q$-form field ($n=0$) and a series of massive $q$-form fields  ($n\geq 1$) on the de Sitter brane.

The Kaluza-Klein reduction procedure can also be applied to the gravity part of the action (\ref{action1}). We can consider the tensor mode of the perturbation of the metric  $g_{MN}$. By implementing the KK decomposition and integrating with respect to the extra dimension, the first term $\int d^5 x \sqrt{-g} \frac{1}{2}R$ reduces to $S_{eff} \propto \int d^4 x \sqrt{-\gamma} \frac{1}{2}R^{(4)}$, where $R^{(4)}$ represents the effective four-dimensional Ricci scalar on the brane.
As for the second term $\int d^5 x \sqrt{-g}\left[ -\frac{1}{2}\nabla_M \phi \nabla^M \phi - V(\phi) \right]$, which describes the chameleon scalar field, it becomes a constant.
	
To recover the four-dimensional effective theory at low energy scale, we have to ensure that the zero modes  with $m_n=0$  can be localized on the brane. The zero modes can be solved from (\ref{q-form schro}) by the factorizing method:
  \begin{eqnarray}
\label{q 0}
 \Upsilon_{0}(z)=N_0 \text{e}^{\frac{(3-2q)\tilde{A}}{2}},
\end{eqnarray}	
where $N_0$ is a normalization constant.
Therefore,  the condition for localization of zero mode is reduced to
\begin{eqnarray}
    \int^{+\infty}_{-\infty} \Upsilon_0^2 (z) dz=1,
\end{eqnarray}
or equivalently,
\begin{eqnarray}
\label{q-form condition1}
 	\int^{+\infty}_{-\infty} \left[a(z)b(\phi)\right]^{3-2q} dz <\infty.
\end{eqnarray}
Obviously, the condition for localization requires $\left[a(z)b(\phi)\right]^{3-2q}_{z\rightarrow\pm\infty}\rightarrow0$. Since the exponentiation $3-2q$ for the scalar and vector fields have opposite signs with that of KR field, which means that when the zero mode of the scalar and vector field localize on the dS brane, the KR filed can not be localized on the dS brane at the same time. 
As KR particles have not been discovered yet on the brane,
it is reasonable for the vector field to be localized but KR field to not be localized on the brane.

In summary, in chameleon gravity, the scalar and vector fields can be localized on the de Sitter brane, while the KR field can not be localized at the same time. The condition to localize both the scalar and vector fields is
\begin{eqnarray}
\label{q-form condition1}
 	\int^{+\infty}_{-\infty} \left[a(z)b(\phi)\right] dz <\infty.
\end{eqnarray}
In addition, since $a(z)$ is an even function and $\phi(z)$ is odd, to ensure that the action  does not vanishes, $b(\phi)$ can not be an odd function of $\phi$.

On the condition that the above condition is satisfied, the effective potentials and the four-dimensional effective actions of the scalar and vector fields can be obtained.
\subsection{Scalar field}
For the scale field $\Phi(x^{\mu},z)$, the KK decomposition (\ref{q-form kk}) turns to
\begin{eqnarray}
\label{scalar kk}
    \Phi(x^{\mu},z) = \sum_n \varphi^{(n)}(x^{\mu})\chi_{n}(z) \text{e}^{-\frac{3}{2}\tilde{A}}.
\end{eqnarray}
and the effective potential for the scalar field reads
\begin{eqnarray}
\label{scalar potential}
    V_{\text{s}} (z) = \frac{3}{2}\tilde{A}'' + \frac{9}{4} \tilde{A}'^2.
\end{eqnarray}
Therefore, the massless scalar zero mode reads
  \begin{eqnarray}
\label{vector 0}
    \chi_{0} (z)\propto \text{e}^{\frac{3\tilde{A}}{2}}.
\end{eqnarray}	
The four-dimensional effective action for the scalar field becomes
\begin{eqnarray}
\label{scalar eff action}
    S_{\text{s}}=-\sum_{n\geq1} \int d^4 x \sqrt{-\gamma}\left(\frac{1}{2}\pt^{\mu} \varphi^{(0)} \pt_{\mu}\varphi^{(0)}+\frac{1}{2}\pt^{\mu} \varphi^{(n)} \pt_{\mu}\varphi^{(n)}+m_{(n)}^2 \varphi^{(n)}\varphi^{(n)}  \right).
\end{eqnarray}
which describes a massless and a series of massive scalar fields,
\subsection{Vector field}
Then we consider the vector field $ A_{\mu}(x, z)$ with the gauge $A_5=0$. The KK decomposition is
\begin{eqnarray}
\label{vector kk}
    A_{\mu}(x, z)= \sum_{n} a^{(n)}_{\mu}(x)\alpha_{n} (z) \text{e}^{-\frac{\tilde{A}}{2}}.
\end{eqnarray}	
The effective potential $V_{\text{v}} (z)$ for the vector field is given by
  \begin{eqnarray}
\label{vector potential}
    V_{\text{v}} (z) = \frac{1}{2}\tilde{A}'' + \frac{1}{4} \tilde{A}'^2,
\end{eqnarray}
and the massless vector zero mode reads
  \begin{eqnarray}
\label{vector 0}
    \alpha_{0} (z)\propto \text{e}^{\frac{\tilde{A}}{2}}.
\end{eqnarray}		
Then, the four-dimensional effective action for the scalar field reads
\begin{eqnarray}
\label{vector eff action}
  S_\text{v}=-\sum_{n\geq1}\int d^4 x \sqrt{-\gamma}\left( \frac{1}{4} f^{(0)\mu\nu}f^{(0)}_{\mu\nu} +\frac{1}{4}f^{(n)\mu\nu}f^{(n)}_{\mu\nu}+\frac{1}{2}m_n^2a^{(n)\mu} a^{(n)}_{\mu} \right),
\end{eqnarray}
which describes a massless vector field and a series of massive vector fields.

\section{Setting up $b(\phi)$}
In the last section, we  demonstrated  that the zero modes of both scalar and vector fields can be localized on a de sitter brane generated by a chameleon scalar field. With the provision of $b(\phi)$, we can further explore additional localized characteristics. In this section, we will study two examples that exhibit novel  properties.
\subsection{Case I: Asymmetric effective potentials}
\begin{figure}[!htb]
    \begin{center}
    \subfigure[The scalar field]{
        \includegraphics[width=7cm]{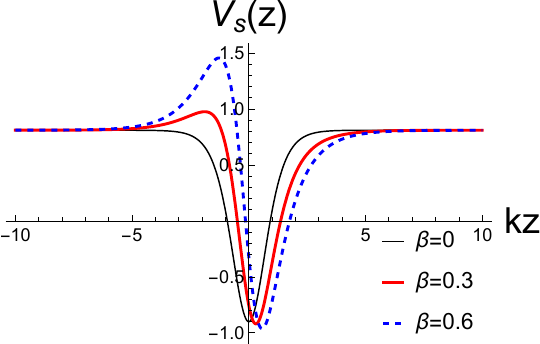}}
    \subfigure[The  vector field]{
        \includegraphics[width=7cm]{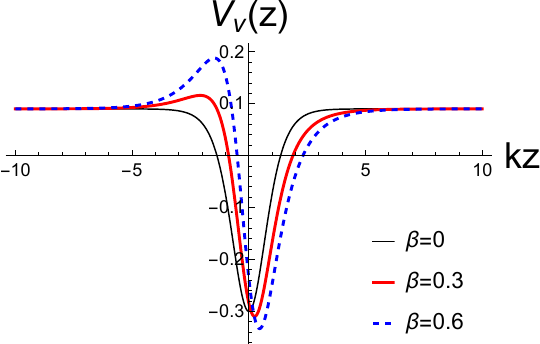}}
    \end{center}
    \caption{The effective potential  of the scalar and vector fields. The parameter $p$ is sect as $p=0.6$.}\label{potential 1}
    \end{figure}

\begin{figure}[!htb]
    \begin{center}
    \subfigure[Scalar field]{
        \includegraphics[width=7cm]{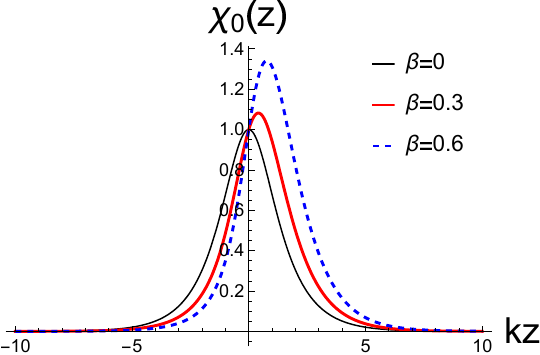}}
    \subfigure[Vector field]{
        \includegraphics[width=7cm]{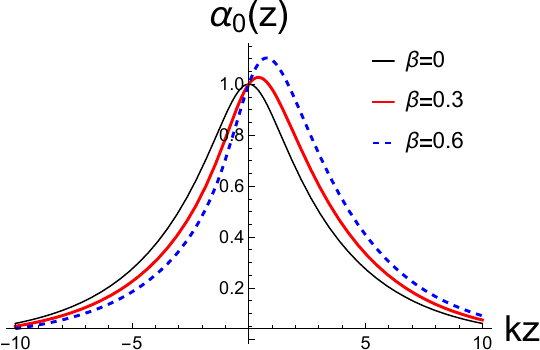}}
    \end{center}
    \caption{The zero modes  of the scalar and vector fields. The parameter $p$ is sect as $p=0.6$.}\label{zero mode 1}
    \end{figure}

For the first case,  we consider the assumption that  $b(\phi)=\text{e}^{\beta \phi}$, which constitutes the initial form of the conformal factor $b(\phi)$ \cite{Waterhouse2006}.  This particular form may stem from processes like dimensional reduction and was widely used in string theory, see Refs. \cite{Damour1994,Fradkin1985,Callan1985,Callan1986} for instance. The gravity dual of
Little String Theory also have similar action \cite{Antoniadis2012}. In this assumption,  since function $b(\phi)$ is asymmetric, while the warp factor $a(z)$ is an even function, the effective potentials and zero modes become asymmetric, which are plotted in Figs.~\ref{potential 1} and \ref{zero mode 1}. When $\beta=0$, our model everts to the standard case in general relativity \cite{Guo2013}.  

Therefore, we can obtain asymmetric effective potentials and the KK modes in a thick brane with $Z_2$ symmetry in chameleon gravity. It was found in Refs.\cite{Liu2009a,Du2013} that the   asymmetry of the effective potential may affects the  spectrum of KK modes.

\subsection{Case II: Volcano-like effective potentials}
In a de Sitter brane in general relativity, the effective potential of matters field is usually P\"ochl–Teller (PT) potentials, and  the zero modes of the scalar field and the vector field are localized on the brane \cite{Liu2009a,Guo2013}. Moreover, it was found that the excited KK mode with a mass of $m_s=\sqrt{2}H\sim 10^{-33}\text{eV}$ for the scalar field is also a bound state \cite{Liu2009a,Guo2013}. Since the mass $m_s$ is small, this excited KK mode of the scalar field may have observable effects and could be inconsistent with experimental observations.

\begin{figure}[!htb]
    \begin{center}
    \subfigure[Scalar field]{
        \includegraphics[width=7cm]{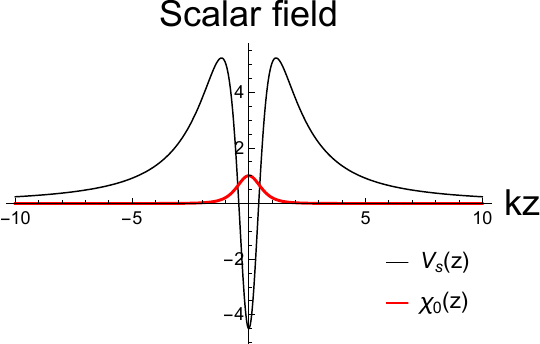}}
    \subfigure[Vector field]{
        \includegraphics[width=7cm]{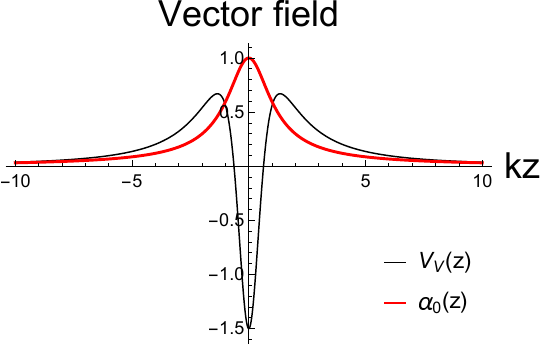}}
    \end{center}
    \caption{The zero modes and effective potentials of  the scalar and vector fields.}\label{case2}
    \end{figure}

In this subsection, we will show that in chameleon gravity, the effective potential could be a volcano-like potential so that all the massive KK modes are not localized. We can assume that $b(\phi)=\frac{\cos^2\left(\frac{\phi}{\phi_0}\right){}^{-p/2}}{\left(\tanh ^{-1}\left(\sin\left(\frac{\phi }{\phi_0}\right)\right){}^2+1\right){}^{3/2}}$.
With this assumption, the effective potential of the scalar and vector fields becomes
 \begin{eqnarray}
\label{scalar potential22}
    V_{\text{s}} (z) &=& \frac{9 k^2 \left(11 k^2 z^2-2\right)}{4 \left(k^2 z^2+1\right)^2},\\
\label{vector potential22}
    V_{\text{v}} (z) &=& \frac{3 k^2 \left(5 k^2 z^2-2\right)}{4 \left(k^2 z^2+1\right)^2},
\end{eqnarray}
which are volcano-like potentials. And  both the scalar field and vector field KK modes have only one bound state, which is the zero mode. The shapes of the effective potentials and zero modes are plotted in Fig.~\ref{case2}.

\section{Conclusion}\label{Conclusion}
In this work, we investigated the localization of $q$-form fields on a de Sitter brane construct by a chameleon scalar field. It was found that the de Sitter brane solution in chameleon gravity is the same as the one in general relativity. The scalar and vector fields can be localized on the de Sitter brane, while the KR field can not be localized at the same time. The condition for localization of the scalar and vector fields is obtained. Furthermore, we investigated the localization characteristics in two examples of given conformal factor $b(\phi)$. In the first case, we assume that $b(\phi)=\text{e}^{\beta \phi}$ , which is inspired by the string theory.
it was shown that the effective potentials and KK modes of the matter fields becomes asymmetric even though the de Sitter brane has $Z_2$ symmetry. In the second case, we generated volcano-like effective potentials in the de Sitter brane by assuming that $b(\phi)=\frac{\cos^2\left(\frac{\phi}{\phi_0}\right){}^{-p/2}}{\left(\tanh ^{-1}\left(\sin\left(\frac{\phi }{\phi_0}\right)\right){}^2+1\right){}^{3/2}}$. In this case, only the zero mode of the scalar and vector fields are localized on the brane.

\acknowledgments
This work was supported by the Natural Science Foundation of Hunan Province, China (Grant No.~2022JJ40033), and the National Natural Science Foundation of China (Grant No.~12305061 and No.~12247101).  


\providecommand{\href}[2]{#2}\begingroup\raggedright\endgroup

\end{document}